\documentclass[runningheads]{llncs}
\usepackage{graphicx}
\usepackage{times}
\usepackage{latexsym}
\usepackage{todonotes}
\usepackage{listings}
\usepackage{subfig}
\usepackage{multirow}
\usepackage{comment}
\usepackage{url}
\usepackage{subfig}
\usepackage[layout=footnote]{fixme}
\graphicspath{ {./img/} }

\begin{document}

\title{Domain-independent Extraction of Scientific Concepts from Research Articles}

\titlerunning{Domain-independent Extraction of Scientific Concepts}
\author{Arthur Brack\orcidID{0000-0002-1428-5348}
\and Jennifer D'Souza\orcidID{0000-0002-6616-9509} 
\and \\ Anett Hoppe\orcidID{0000-0002-1452-9509}
\and S\"oren Auer\orcidID{0000-0002-0698-2864} 
\and \\ Ralph Ewerth\orcidID{0000-0003-0918-6297}}

\authorrunning{A. Brack et al.}
%
\institute{Leibniz Information Centre for Science and Technology (TIB), Hanover, Germany\\
\email{\{arthur.brack|jennifer.dsouza|anett.hoppe| \\
soeren.auer|ralph.ewerth\}@tib.eu}}

\maketitle              
\begin{abstract}
We examine the novel task of 
\textit{domain-independent scientific concept extraction from abstracts of scholarly articles} and present two contributions. 
First, we suggest a set of generic scientific concepts that have been identified in a systematic annotation process. 
This set of concepts is utilised to annotate a corpus of scientific abstracts from 10 domains of Science, Technology and Medicine at the phrasal level in a joint effort with domain experts. 
The resulting dataset is used in a set of benchmark experiments to (a) provide baseline performance for this task, (b) examine the transferability of concepts between domains.
Second, we present two deep learning systems as baselines. 
In particular, we propose active learning to deal with different domains in our task.
The experimental results show that (1) a substantial agreement is achievable by non-experts after consultation with domain experts, (2) the baseline system achieves a fairly high F1 score, (3) active learning enables us to nearly halve the amount of required training data.


\end{abstract}

\section{Introduction}
Scholarly communication as of today is a document-centric process. Research results are usually conveyed in written articles, as a PDF file with text, tables and figures. Automatic indexing of these texts is limited and generally does not access their semantic content. There are thus severe limitations how current research infrastructures can support scientists in their work: finding relevant research works, comparing them, and compiling summaries is still a tedious and error-prone manual work. The strong increase in the number of published research papers aggravates this situation~\cite{bornmann15growth}. 

Knowledge graphs are recognised as an effective approach to facilitate semantic search~\cite{Balog2018EntityOrientedS}. 
For academic search engines, Xiong et al.~\cite{Xiong2017ExplicitSR} have shown that exploiting knowledge bases like Freebase can improve search results. 
However, the introduction of new scientific concepts occurs at a faster pace than knowledge base curation, resulting in a large gap in knowledge base coverage of scientific entities~\cite{Ammar2018ConstructionOT}, 
e.g. the task \textit{geolocation estimation of photos} from the Computer Vision field is neither present in Wikipedia nor in more specialised knowledge bases like Computer Science Ontology (CSO)~\cite{Salatino2018TheCS} or ``Papers with code''~\cite{PWC}.
Information extraction from text helps to identify emerging entities and to populate knowledge graphs~\cite{Balog2018EntityOrientedS}.
Thus, information extraction from scientific texts is a first vital step towards a fine-grained research knowledge graph in which research articles are described and interconnected through entities like tasks, materials, and methods. 
Our work is motivated by the idea of the automatic construction of a research knowledge graph.

Information extraction from scientific texts, obviously, differs from its general domain counterpart: 
Understanding a research paper and determining its most important statements demands certain expertise in the article's domain. 
Every domain is characterised by its specific terminology and phrasing which is hard to grasp for a non-expert reader.
In consequence, extraction of scientific concepts from text would entail the involvement of domain experts and a specific design of an extraction methodology for each scientific discipline -- both requirements are rather time-consuming and costly.

At present, a structured study of these assumptions is missing. 
We thus present the task of \textit{domain-independent scientific concept extraction}. 
This article examines the intuition that most domain-specific articles share certain core concepts such as the mentions of research tasks, used materials, or data. 
If so, these would allow a domain-independent information extraction system, which does not reach all semantic depths of the analysed article, but still provides some science-specific structure.

In this paper, we introduce a set of science concepts that generalise well over the set of examined domains (10 disciplines from Science, Technology and Medicine (STM)). These concepts have been identified in a systematic, joint effort of domain experts and non-domain experts. The inter-coder agreement is measured to ensure the adequacy and quality of concepts. A set of research abstracts has been annotated using these concepts and the results are discussed with experts from the corresponding fields. The resulting dataset serves as a basis to train two baseline deep learning classifiers. In particular, we present an active learning approach to reduce the number of required training data. The systems are evaluated in different experimental setups.

Our main contributions can be summarised as follows:
(1) We introduce the novel task \emph{domain-independent scientific concept extraction}, which aims at automatically extracting scientific entities in a domain-independent manner.
(2) We release a new corpus that comprises 110 abstracts of 10 STM domains annotated at the phrasal level. 
Additionally, we release a silver-labelled corpus with 62K automatically annotated abstracts of Elsevier with CCBY license and 1.2 Mio. extracted unique concepts comprising 24 domains.
(3) We present two baseline deep learning systems for this task, including an active learning approach.
To the best of our knowledge, this is the first approach that applies active learning to scholarly texts. We demonstrate that about half of the training data are sufficient to maintain the performance when using the entire training set. 
(4)~We make our corpora and source code publicly available to facilitate further research.

\section{Related Work}
This section gives a brief overview of existing annotated scientific corpora before some exemplary applications for domain-independent information extraction from scientific papers and the respective state of the art are introduced.

\subsection{Scientific corpora}
\label{sec:corpora}
\textbf{Sentence level annotation.}
Early approaches for semantic structuring of research papers focused on sentences as the basic unit of analysis. 
This enables, for instance, automatic highlighting of relevant paper passages to enable efficient assessment regarding quality and relevance. 
Several ontologies have been created that focus on the rhetorical~\cite{Groza2006SALTSA,Constantin2016TheDC}, argumentative~\cite{teufel2009towards,Liakata2010CorporaFT} or 
activity-based \cite{pertsas2017scholarly} structure of research papers.

Annotated datasets exist for several domains, e.g. PubMed200k~\cite{Dernoncourt2017PubMed2R} from biomedical randomized controlled trials, NICTA-PIBOSO~\cite{Kim2011AutomaticCO} from evidence-based medicine, Dr. Inventor~\cite{Fisas2015OnTD} from Computer Graphics, Core Scientific Concepts (CoreSC) \cite{Liakata2010CorporaFT} from Chemistry and Biochemistry, and Argumentative Zoning (AZ)~\cite{teufel2009towards} from Chemistry and Computational Linguistics, Sentence Corpus~\cite{Chambers2013} from Biology, Machine Learning and Psychology. Most datasets cover only a single domain, while few other datasets cover three domains. Several machine learning methods have been proposed for scientific sentence classification~\cite{Jin2018HierarchicalNN,Dernoncourt2017PubMed2R,Fisas2015OnTD,liakata2012automatic}.

\textbf{Phrase level annotation.}
More recent corpora have been annotated at phrasal level.
SciCite\cite{Cohan2019StructuralSF} and ACL~ARC~\cite{Jurgens2018MeasuringTE} are datasets for citation intent classification from Computer Science, Medicine, and Computational Linguistics.
ACL~RD-TEC~\cite{handschuh2014acl} from Computational Linguistics aims at extracting scientific technology and non-technology terms.
ScienceIE17~\cite{augenstein2017semeval} from Computer Science, Material Sciences, and Physics contains three concepts \textsc{Process}, \textsc{Task} and \textsc{Material}.
SciERC~\cite{Luan2018MultiTaskIO} from the machine learning domain contains six concepts \textsc{Task}, \textsc{Method}, \textsc{Metric}, \textsc{Material}, \textsc{Other-Scien\-tific\-Term} and \textsc{Generic}. 
Each corpus covers at most three domains.

\textbf{Experts vs. non-experts.}
The aforementioned datasets were usually annotated by domain experts~\cite{Dernoncourt2017PubMed2R,Kim2011AutomaticCO,augenstein2017semeval,Luan2018MultiTaskIO,handschuh2014acl,Liakata2010CorporaFT}. 
In contrast, Teufel et al.~\cite{teufel2009towards} explicitly use non-experts in their annotation tasks, arguing that text understanding systems can use general, rhetorical and logical aspects also when qualifying scientific text. According to this line of thought, more researchers used (presumably cheaper) non-expert annotation as an alternative~\cite{Fisas2015OnTD,Chambers2013}.

Snow~et.~al.~\cite{Snow2008CheapAF} provide a study on expert versus non-expert performance for general, non-scientific annotation tasks. They state that about four non-experts (Mechanical Turk workers, in their case) were needed to rival the experts' annotation quality. However, systems trained on data generated by non-experts showed to benefit from annotation diversity and to suffer less from annotator bias. A recent study \cite{PustuIren2019InvestigatingCO} examines the agreement between experts and non-experts for visual concept classification and person recognition in historical video data. For the task of face recognition, training with expert annotations lead to an increase of only 1.5 \% in classification accuracy.

\textbf{Active learning in Natural Language Processing (NLP).}
To the best of our knowledge, active learning has not been applied to classification tasks for scientific text yet.
Recent publications demonstrate the effectiveness of active learning for NLP tasks such as \textit{Named Entity Recognition} (NER)~\cite{Shen2017DeepAL} and sentence classification~ \cite{Zhang2016ActiveDT}. Siddhant~and~Lipton~\cite{Siddhant2018DeepBA} and Shen~et.~al.~\cite{Shen2017DeepAL} compare several sampling strategies on NLP tasks and show that \emph{Maximum Normalized Log-Probability} (MNLP) based on uncertainty sampling performs well in NER. 
 
\subsection{Applications for domain-independent scientific information extraction}
\label{sec:applications}
\textbf{Academic search engines.}
Academic search engines  such as Google Scholar~\cite{GoogleScholar}, Microsoft Academic~\cite{MicrosoftAcademic} and Semantic Scholar~\cite{SemanticScholar}
specialise in search of scholarly literature. 
They exploit graph structures such as the Microsoft Academic Knowledge Graph~\cite{MicrosoftAcademicKG}, SciGraph~\cite{SpringerSciGraph}, or the Semantic Scholar Corpus~\cite{Ammar2018ConstructionOT}.
These graphs interlink the papers through meta-data such as citations, authors, venues, and keywords, but not through deep semantic representation of the articles' content.

However, first attempts towards a more semantic representation of article content exist: Ammar et al.~\cite{Ammar2018ConstructionOT} interlink the Semantic Scholar Corpus with DBpedia~\cite{Lehmann2015DBpediaA} and Unified Medical Language System (UMLS)~\cite{Bodenreider2004TheUM} using entity linking techniques.
Yaman et al.~\cite{Yaman2019InterlinkingSA} connect SciGraph with DBpedia person entities. 
Xiong et al.~\cite{Xiong2017ExplicitSR} demonstrate that academic search engines can greatly benefit from exploiting general-purpose knowledge bases. However, the coverage of science-specific concepts is rather low \cite{Ammar2018ConstructionOT}. 

\textbf{Research paper recommendation systems.}
Beel et al.~\cite{Beel2015ResearchpaperRS} provide a comprehensive survey about research paper recommendation systems. Such systems usually
employ different strategies (e.g. content-based and collaborative filtering) and several data sources (e.g. text in the documents, ratings, feedback, stereotyping).
Graph-based systems, in particular, exploit citation graphs and genes mentioned in the papers~\cite{Lao2010RelationalRU}. 
Beel et al. conclude that it is not possible to determine the most
effective recommendation approach at the moment. However, we believe that a fine-grained research knowledge graph can improve such systems. Although ``Papers with code''~\cite{PWC}  is not a typical recommendation system, it allows researchers to browse easily for papers from the field of machine learning that address a certain task.

\section{Domain-independent scientific concept extraction: A corpus}
In this section, we introduce the novel task of \textit{domain-independent extraction of scientific concepts} and present an annotated corpus. 
As the discussion of related work reveals, the annotation of scientific resources is not a novel task. However, most researchers focus on at most three scientific disciplines and on expert-level annotations. 
In this work, we explore the domain-independent annotation of scientific concepts based on abstracts from ten different science domains. Since other studies have also shown that non-expert annotations are feasible for the general and scientific domain, we go for a cost-efficient middle course: annotations of non-experts experienced in the annotation task and consultation with domain-experts.
Finally, we explore how well state-of-the-art machine learning approaches do perform on this novel, domain-independent information extraction task and whether active learning can save annotation costs.
The base corpus, which we make publicly available, and the annotation process are described below. 

\subsection{OA STM Corpus}
\label{sec:stm_corpus}
The OA STM corpus~\cite{OASTM} is a set of open access (OA) articles from various domains in Science, Technology and Medicine (STM). It was published in 2017 as a platform for benchmarking methods in scholarly article processing, amongst other scientific information extraction. The dataset contains a selection of 110 articles from 10 domains, namely Agriculture (\textit{Agr}), Astronomy (\textit{Ast}), Biology (\textit{Bio}), Chemistry (\textit{Che}), Computer Science (\textit{CS}), Earth Science (\textit{ES}), Engineering (\textit{Eng}), Materials Science (\textit{MS}), Mathematics (\textit{Mat}), and Medicine (\textit{Med}).
While the original corpus contains full articles, this first annotation cycle focuses on the articles' abstracts.

\subsection{Annotation process}
The OA STM Corpus is used as a base for (a) the identification of potential domain-independent concepts; (b) a first annotated corpus for baseline classification experiments. 
Main actors in the annotation process were two post-doctoral researchers with a background in computer science (acting as non-expert annotators); their basic annotation assumptions were checked by experts from the respective domains. 

\begin{table}[htb]
\small
\caption{The four core scientific concepts that were derived in this study}
\begin{tabular}{|p{12cm}|}
\textsc{\textbf{Process}} Natural phenomenon or activities, e.g. growing (\textit{Bio}), reduction (\textit{Mat}), flooding (\textit{ES}).                                          \\
\textsc{\textbf{Method}} A commonly used procedure that acts on entities, e.g. powder X-ray (\textit{Che}), the PRAM analysis (\textit{CS}), magnetoencephalography (\textit{Med}).                                  \\
\textsc{\textbf{Material}} A physical or abstract entity used in scientific experiments or proofs, e.g. soil (\textit{Agr}), the moon (\textit{Ast}), the carbonator (\textit{Che}).                                                   \\
\textsc{\textbf{Data}} The data themselves, measurements, or quantitative or qualitative characteristics of entities, e.g. rotational energy (\textit{Eng}), tensile strength (\textit{MS}), 3D time-lapse seismic data (\textit{ES}).
\end{tabular}
\label{table:0}
\end{table}

\textbf{Pre-annotation.} A literature review of annotation schemes~\cite{Liakata2010CorporaFT,augenstein2017semeval,liakata2012automatic,Constantin2016TheDC} provided a seed set of potential candidate concepts. 
Both non-experts independently annotated a subset of the STM abstracts with these concepts and discussed the outcome. In a three-step process, the concept set was pruned to only contain those which seemed suitably transferable between domains. Our set of \textit{generic} scientific concepts consists of \textsc{Process}, \textsc{Method}, \textsc{Material}, and \textsc{Data} (see Table~\ref{table:0} for their definitions). We also identified \textsc{Task}~\cite{augenstein2017semeval}, \textsc{Object}~\cite{liakata2012automatic}, and \textsc{Results}~\cite{Constantin2016TheDC}, however, in this study we do not consider nested span concepts, hence we leave them out since they were almost always nested with the other scientific entities (e.g. a \textsc{Result} may be nested with \textsc{Data}).

\textbf{Phase I.} Five abstracts per domain (i.e. 50 abstracts) were annotated by both annotators and the inter-annotator agreement was computed using Cohen's $\kappa$ \cite{cohen1960coefficient}.
Results showed a moderate inter-annotator agreement of 0.52 $\kappa$. 

\textbf{Phase II.} The annotations were then presented to subject specialists who each reviewed (a) the choice of concepts and (b) annotation decisions on the respective domain corpus. The interviews mostly confirmed the concept candidates as generally applicable. 
The experts' feedback on the annotation was even more valuable: The comments allowed for a more precise reformulation of the annotation guidelines, including illustrating examples from the corpus. 

\textbf{Consolidation. }
Finally, the 50 abstracts from phase I were reannotated by the non-experts. 
Based on the revised annotation guidelines,  
a substantial agreement of 0.76~$\kappa$ could be reached (see Table~\ref{table:2}).
Subsequently, the remaining 60 abstracts (six per domain) were annotated by one annotator. 
This last phase also involved reconciliation of the previously annotated 50 abstracts to obtain a gold standard corpus.

\begin{table}[htb]
\centering
\caption{Per-domain and overall inter-annotator agreement (Cohen's Kappa $\kappa$) for \textsc{Process}, \textsc{Method}, \textsc{Material}, and \textsc{Method} scientific concept annotation}
\begin{tabular}{l|rrrrrrrrrr|r}
	& \textit{Med}	&\textit{MS}	&\textit{CS}	&\textit{ES}	&\textit{Eng}	&\textit{Che}	&\textit{Bio}	&\textit{Agr}	&\textit{Mat}	&\textit{Ast} &\textit{Overall} \\ \hline
$\kappa$	& 0.94 & 0.90 & 0.85 & 0.81 & 0.79 & 0.77 & 0.75 & 0.60 & 0.58 & 0.57 & 0.76 
\end{tabular}
\label{table:2}
\end{table}

\subsection{Corpus characteristics}
Table~\ref{table:1} shows some characteristics of the resulting corpus. 
The corpus has a total of 6,127 scientific entities, including 2,112 \textsc{Process}, 258 \textsc{Method}, 2,099 \textsc{Material}, and 1,658 \textsc{Data} concept entities. The number of entities per abstract in our corpus directly correlates with the length of the abstracts (Pearson's \textit{R} 0.97). 
Among the concepts, \textsc{Process} and \textsc{Material} directly correlate with abstract length (\textit{R} 0.8 and 0.83, respectively), while \textsc{Data} has only a slight correlation (\textit{R} 0.35) and \textsc{Method} has no correlation (\textit{R} 0.02). 
The domains \textit{Bio}, \textit{CS}, \textit{Ast}, and \textit{Eng} contain the most of \textsc{Process}, \textsc{Method}, \textsc{Material}, and \textsc{Data} concepts, respectively.

\begin{table*}[htb]
\centering
\small
\caption{The annotated corpus characteristics in terms of size and the number of scientific concept phrases}
\begin{tabular}{c|rrrrrrrrrr}
                         & \textit{Ast} & \textit{Agr} & \textit{Eng} & \textit{ES} & \textit{Bio} & \textit{Med} & \textit{MS} & \textit{CS} & \textit{Che} & \textit{Mat} \\ \hline
Avg. \# Tokens/Abstract                      & 382                           & 333                           & 303                           & 321                          & 273                           & 274                           & 282                          & 253                          & 217                           & 140                           \\
\# Gold scientific concept phrases                     & 791                           & 741                           & 741                           & 698                          & 649                           & 600                           & 574                          & 553                          & 483                           & 297                           \\
\# Unique gold scientific concept phrases        & 663                           & 631                           & 618                           & 633                          & 511                           & 518                           & 493                          & 482                          & 444                           & 287                           \\
\# \textsc{Process}  & 241                           & 252                           & 248                           & 243                          & 281                           & 244                           & 178                          & 220                          & 149                           & 56                            \\

\# \textsc{Method}     & 19                            & 28                            & 27                            & 9                            & 15                            & 33                            & 27                           & 66                           & 27                            & 7                             \\

\# \textsc{Material} & 296                           & 292                           & 208                           & 249                          & 291                           & 191                           & 231                          & 102                          & 188                           & 51                            \\
\# \textsc{Data}   & 235                           & 169                           & 258                           & 197                          & 62                            & 132                           & 138                          & 165                          & 119                           & 183  
\end{tabular}
\label{table:1}
\end{table*}

\section{Experimental setup: Two baseline classifiers}
\label{sec:benchmark}

The current state-of-the-art 
for scientific entity extraction is Beltagy et al.'s system~\cite{Beltagy2019SciBERTPC}. We use their NER task-specific deep learning architecture atop SciBERT embeddings with a Conditional Random Field (CRF) based sequence tag decoder~\cite{Ma2016EndtoendSL} and BILOU (beginning, inside, last, outside, unit) tagging scheme.
The following classifiers are implemented in AllenNLP~\cite{gardner2018allennlp}. 
We report span-based micro-averaged F1 scores and use the ScienceIE17~\cite{augenstein2017semeval} evaluation script.

\subsection{Traditionally trained classifiers}

Using the above mentioned architecture, we train one model with data from all domains combined. We refer to this model as the \textit{domain-independent} classifier. Similarly, we train 10 models for each domain in our corpus -- the \textit{domain-specific} classifier.

To obtain a robust evaluation of models, we perform five-fold cross-validation experiments. In each fold experiment, we train a model on 8 abstracts per domain (i.e. 80 abstracts), tune hyperparameters on 1 abstract per domain (i.e. 10 abstracts), and test on the remaining 2 abstracts per domain (i.e. 20 abstracts) ensuring that the data splits are not identical between the folds. 
All results reported in the paper are averaged over the five folds. Please note that 8 abstracts have about 445 concepts so that the training data should be sufficient for the domain-dependent classifier.

\subsection{Active learning trained classifier}
Based on the results of the aforementioned comparison studies~\cite{Siddhant2018DeepBA,Zhang2016ActiveDT}, we decide to use MNLP~\cite{Shen2017DeepAL} as the sampling strategy in the active learning setting. It is chosen over other possibly suitable candidates such as \emph{Bayesian Active Learning by Disagreement} (BALD)~\cite{Houlsby2011BayesianAL}, which is another powerful strategy, but has higher computational requirements.
The objective involves strategically selecting sentences from the overall dataset in each iteration of the algorithm greedily, aiming at getting greater performance with a minimum number of sentences. In our experiments, we found that adding 4\% of the data to be the most discriminative selection of classifier performance. Therefore, we run 25 iterations of active learning in each stage adding 4\% training data. To obtain a robust evaluation of models, we repeat the experiment for five folds and average the results. 
The models use the same hyperparameters as for the domain-independent classifier. We retrain the model within each iteration and fold. 

\section{Experimental results and discussion}
This section describes the results of the experimental setup and the correlation analysis between inter-annotator agreement and performance of the several classifiers.

\subsection{Traditionally trained classifiers}
Table~\ref{table:4} shows an overview of the \textit{domain-independent} classifier results. The system achieves an $F1$ score of 65.5 ($\pm$ 1.26) in the overall task. For this classifier, \textsc{Material} was the easiest concept with an $F1$ of 71 ($\pm$ 1.88), whereas \textsc{Method} was the hardest concept with an $F1$ of 43 ($\pm$ 6.30). 
The concept \textsc{Method} is also the most underrepresented one in our corpus, which partly explains the poor extraction performance.

\begin{table}[htb]
\small
\caption{The \textit{domain-independent} classifier results in terms of Precision ($P$), Recall ($R$), and F1-score on scientific concepts, respectively, and \textit{Overall}}
\begin{tabular}{p{3.5mm}|rrrr|r}
            & \multicolumn{1}{p{20mm}}{\textsc{Process}} & \multicolumn{1}{p{20mm}}{\textsc{Method}} & \multicolumn{1}{p{20mm}}{\textsc{Material}} & \multicolumn{1}{p{20mm}}{\textsc{Data}}
            & \multicolumn{1}{p{20mm}}{\textit{Overall}} \\ \hline
$P$  & 65.5 ($\pm$ 4.22)   & 45.8 ($\pm$ 13.50)    & 69.2 ($\pm$ 3.55)      & 60.3 ($\pm 4.14$)      & 64.3 ($\pm$ 1.73)   \\
$R$  & 68.3 ($\pm$ 1.93)  & 44.1 ($\pm$ 8.73)   & 73.2 ($\pm$ 4.27)     & 60.0 ($\pm$ 4.84)    & 66.7  ($\pm$ 0.92)    \\
$F1$ & 66.8  ($\pm$ 2.07)  & 43.0 ($\pm$ 6.30)  & 71.0 ($\pm$ 1.88)      & 59.8 ($\pm$ 1.75)      & 65.5 ($\pm$ 1.26)      
\end{tabular}
\label{table:4}
\end{table}

\begin{figure}[htb]
    \center{\includegraphics[width=\linewidth]
        {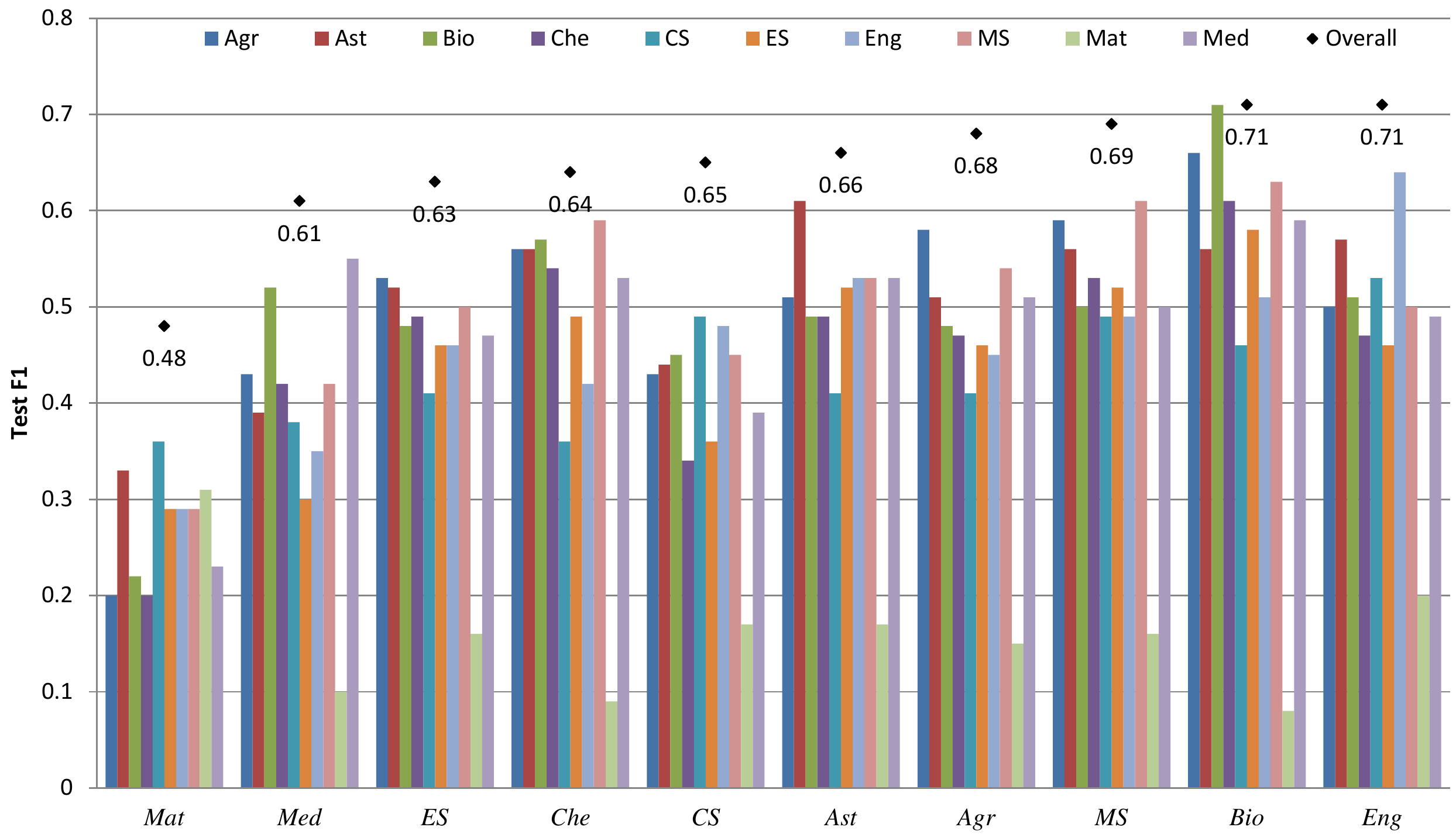}}
    \caption{$F1$ per domain of the 10 \textit{domain-specific} classifiers (as bar plots) and of the \textit{domain-independent} classifier (as scatter plots) for scientific concept extraction; the x-axis represents the 10 test domains}
    \label{fig:2}
\end{figure}

Next, we compare and contrast the 10 \textit{domain-specific}
classifiers according to their capability to extract the concepts from their own domains and in other domains.
The results are shown as $F1$ scores in Figure~\ref{fig:2} where the x-axis represents the 10 test domains. We discuss some observations in the sequel.

\begin{figure}[!htb]
    \centering
    \subfloat[]{\includegraphics[width=0.45\linewidth]
        {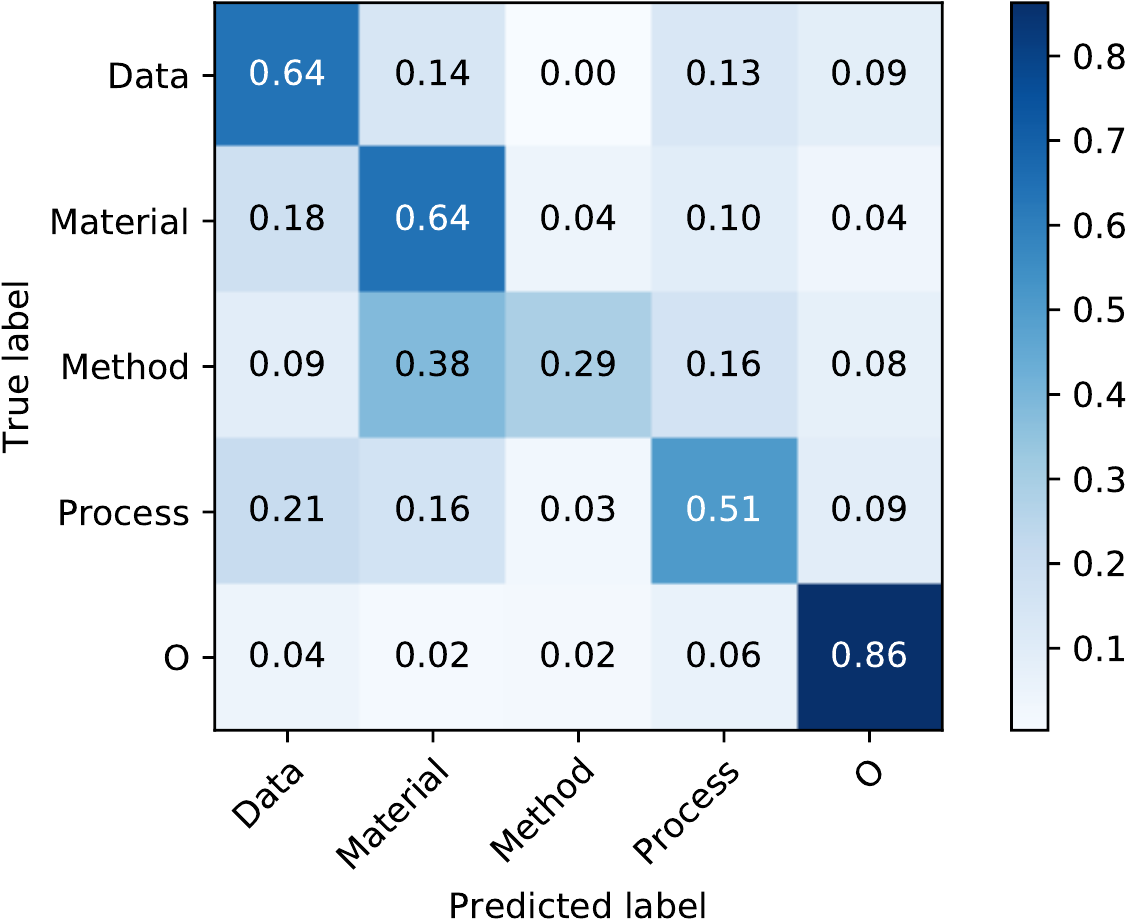}}
    \qquad
    \subfloat[]{\includegraphics[width=0.45\linewidth]
        {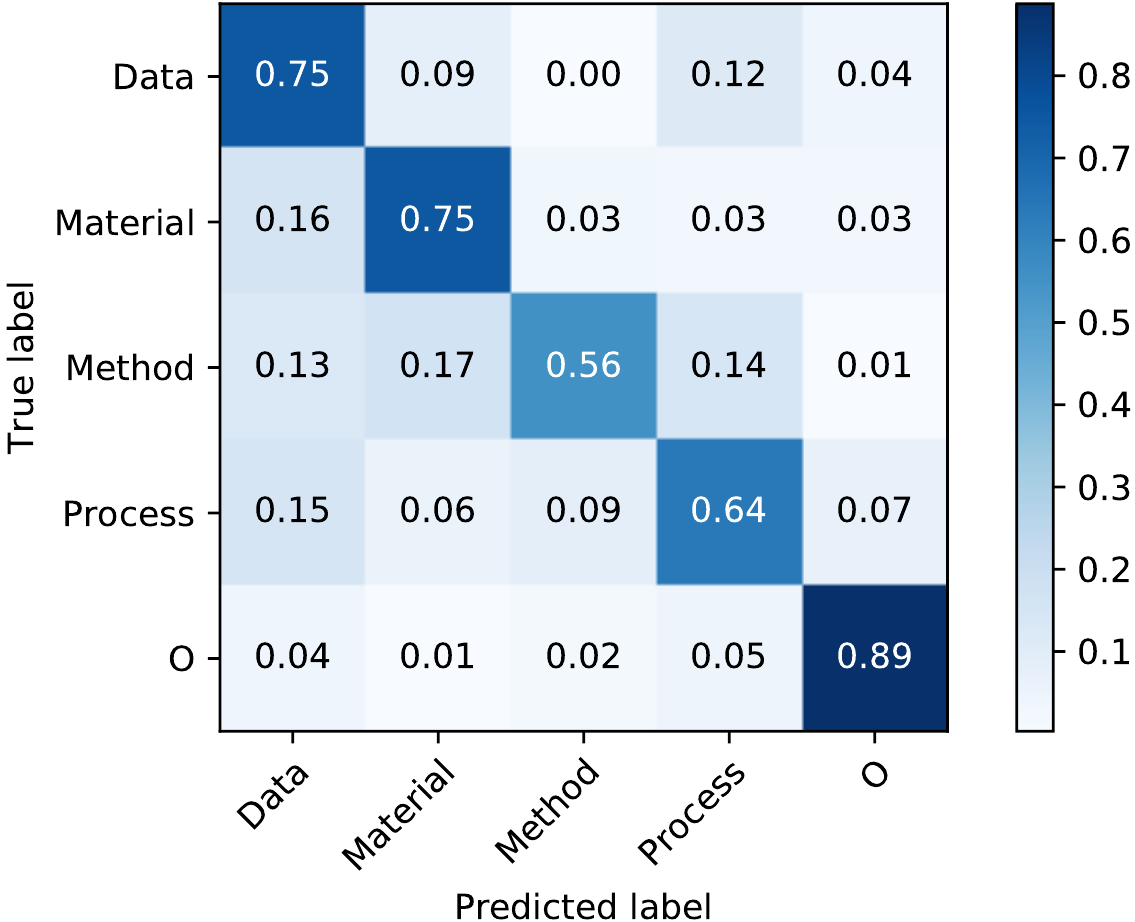}}        
        
    \caption{Confusion matrix for (a) the \textit{CS} classifier and (b) \textit{domain-independent classifier} on \textit{CS} domain predicting concept-type of tokens}
    \label{fig:cm}
\end{figure}

\begin{figure}[htb]
    \center{\includegraphics[width=\linewidth]
        {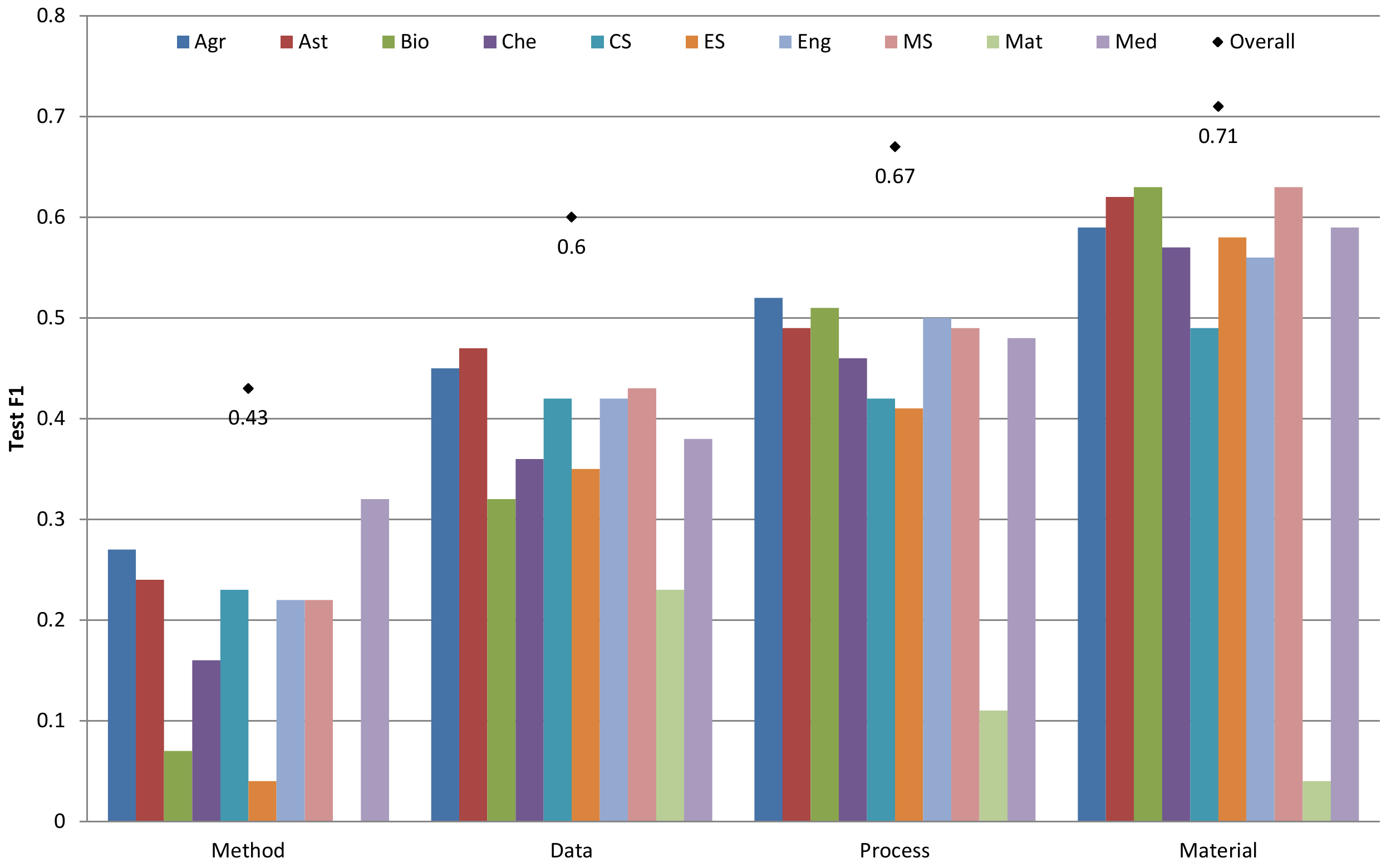}}
    \caption{$F1$ of the 10 \textit{domain-specific} classifier (as bar plots) and the \textit{domain-independent} classifier (as scatter plots) for extracting each scientific concept; the x-axis represents the evaluated concept }
    \label{fig:3}
\end{figure}

\textbf{Most robust domain.} \textit{Bio} (third bar in each domain in Figure~\ref{fig:2}) extracts scientific entities from its own domain at the same performance as the \textit{domain-independent} classifier with an $F1$ score of 71 ($\pm$ 9.0) demonstrating a robust domain. It comprises only 11\% of the overall data, yet the \textit{domain-independent} classifier trained on all data does not outperform it.

\textbf{Most generic domain.} \textit{MS} (the third last bar in each domain in Figure~\ref{fig:2}) exhibits a high degree of domain independence since it is among the top 3 classifiers for seven of the 10 domains (viz. \textit{ES}, \textit{Che}, \textit{CS}, \textit{Ast}, \textit{Agr}, \textit{MS}, and \textit{Bio}).

\textbf{Most specialised domain.} \textit{Mat} (the second last bar in each domain in Figure~\ref{fig:2}) shows the lowest performance in extracting scientific concepts from all domains except itself. Hence it shows to be the most specialised domain in our corpus. Notably, a characteristic feature of this domain is that it has short abstracts (nearly a third of the size of the longest abstracts), so it is also the most underrepresented in our corpus. Also, distinct from the other domains, \textit{Mat} has triple the number of \textsc{Data} entities compared to each of its other concepts, where in the other domains \textsc{Process} and \textsc{Material} are consistently predominant.

\textbf{Medical and Life Science domains.}
The \textit{Med}, \textit{Agr}, and \textit{Bio} domains
show strong domain relatedness. Their respective \textit{domain-specific} classifiers show top five system performances among the three domains, when applied to another domain. For instance, the \textit{Med} domain shows the strongest domain relatedness and is classified best by \textit{Med} (last bar), followed by \textit{Bio} (third bar) and \textit{Agr} (first bar).

\textbf{Domain-independent vs. domain-dependent classifier.}
Except for \textit{Bio} the \textit{domain-independent} classifier clearly outperforms the \textit{domain-dependent} one extracting concepts from their respective domains. To analyse the reason, we investigate the improvements in \textit{CS} domain. We have chosen \textit{CS} exemplary as the size of the domain is slightly below the average and this domain strongly benefits from the \textit{domain-independent} classifier and improves the $F1$ score for the \textit{CS} classifier from 49.5 ($\pm$ 4.22) to 65.9~($\pm$ 1.21). The $F1$ score for span-detection is improved from 73.4 ($\pm$ 3.45) to 82.0 ($\pm$ 3.98). Span-detection usually requires less domain-dependent signals, thus the \textit{domain-independent} classifier can benefit from other domains. 
Accuracy on token-level also improves from 67.7 ($\pm$ 5.35) to 77.5 ($\pm$ 4.42) $F1$, that is correct labelling of the tokens also benefits from other domains. This is also supported by the results in the confusion matrix depicted in Figure \ref{fig:cm} for the \textit{CS} and the \textit{domain-independent} classifier on token-level. 

\textbf{Scientific concept extraction.} 
Figure~\ref{fig:3} depicts the 10 \textit{domain-specific} classifier results for extracting each of the four scientific concepts.
It can be observed that \textit{Agr}, \textit{Med}, \textit{Bio}, and \textit{Ast} classifiers are the best in extracting \textsc{Process}, \textsc{Method}, \textsc{Material}, and \textsc{Data}, respectively.

\subsection{Active learning trained classifier}

\begin{figure}[htb]
    \center{\includegraphics[width=\linewidth]
        {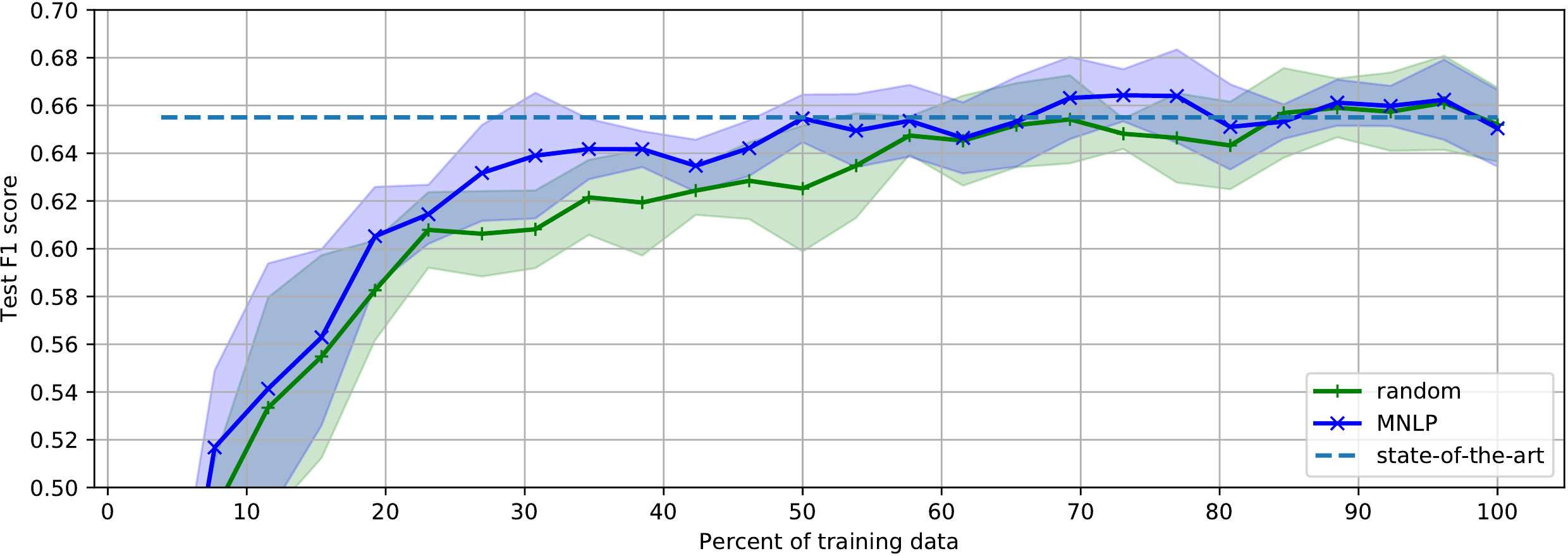}}
    \caption{Progress of active learning with MNLP and random sampling strategy; the areas represent the standard deviation (std) of the F1 score across 5 folds for MNLP and random sampling strategy, respectively}
    \label{fig:al_sen_based}    
\end{figure}

Figure \ref{fig:al_sen_based} shows the results of the active learning experiment.
Table \ref{table:al_results} depicts the results for the fraction of training data when the performance using the entire training dataset is achieved. MNLP clearly outperforms the random baseline. While using only 52 \% of the training data, the best result of the \textit{domain-independent} classifier trained with all training data is surpassed with an $F1$ score of 65.5 ($\pm$ 1.0). For comparison: the random baseline achieves an $F1$ score of 
62.5 ($\pm$ 2.6) with 52 \% of the training data. 
When 76 \% of the data are sampled by MNLP, the best active learning performance across all steps is achieved with an 
$F1$ score of 69.0 on the validation set, having the best $F1$ of
66.4 ($\pm$ 2.0) on the test set. 
For SciERC~\cite{Luan2018MultiTaskIO} and ScienceIE17~\cite{augenstein2017semeval} similar results are demonstrating that MNLP can significantly reduce the amount of labelled data.

\begin{table}[htb]
\caption{Performance of active learning with MNLP and random sampling strategy for the fraction of training data when the performance with entire training dataset is achieved, for SciERC and ScienceIE17 results are reported across 5 random restarts}
\begin{tabular}{l|rrrr}
& \multicolumn{1}{p{20mm}}{training data} & \multicolumn{1}{p{20mm}}{F1 (MNLP)}        & \multicolumn{1}{p{20mm}}{F1 (random)}        & \multicolumn{1}{p{20mm}}{F1 (full data)}    \\ \hline
STM (our corpus)         & 52 \%                    & 65.5 ($\pm$ 1.0) & 62.5 ($\pm$ 2.6) & 65.5 ($\pm$ 1.3) \\
SciERC~\cite{Luan2018MultiTaskIO}      & 62 \%                   & 65.3 ($\pm$ 1.5) & 62.3 ($\pm$ 1.5) & 65.6 ($\pm$ 1.0) \\
ScienceIE17~\cite{augenstein2017semeval} & 38 \%                   & 43.9 ($\pm$ 1.2) & 42.2 ($\pm$ 1.8) & 43.8 ($\pm$ 1.0)
\end{tabular}
\label{table:al_results}
\end{table}

To find out which mix of training data produces the most generic model, we analyse the distribution of sentences in the training data sampled by MNLP. 
As expected, the random sampling strategy uniformly samples sentences from all domains in each iteration.
However, (\textit{Math}, \textit{CS}) are the most and (\textit{Eng}, \textit{MS}) the least preferred domains by MNLP. 
When using 52 \% of the training data, 65.4\% of \textit{Math}, 66.2\% of \textit{CS} sentences were sampled, but only 41.6\% of \textit{Eng} and 37.3\% of \textit{MS}. Thereby all domains are present, that is a heterogeneous mix of sentences sampled by MNLP yields the most generic model with less training data.

\subsection{Correlations between inter-annotator agreement and performance}

In this section, we analyse the correlations of inter-annotator agreement $\kappa$ and the number of annotated concepts per domain (\#) on the performance and variance of the classifiers employing Pearson's correlation coefficient (Pearson's \textit{R}).

\begin{table}[htb]
\caption{Inter-annotator agreement ($\kappa$) and the number of concept phrases (\#) per domain; F1 and std of domain-dependent classifiers on their domains; F1 and std of domain-independent and AL-trained classifier on each domain; the right side depicts correlation coefficients (\textit{R}) of each row with $\kappa$ and the number of concept phrases}
\begin{tabular}{l|rrrrrrrrrr|rr}
  & \textit{Agr} & \textit{Ast} & \textit{Bio} & \textit{Che} & \textit{CS} & \textit{ES} & \textit{Eng} & \textit{MS} & \textit{Mat} & \textit{Med} & \textit{R} $\kappa$ & \textit{R} \# \\ \hline        
inter-annotator agreement ($\kappa$) &         0.6           & 0.57          & 0.75         & 0.77          & 0.85          & 0.81          & 0.79          & 0.9           & 0.58          & 0.94 & 1.00 & -0.02          \\
\# concept phrases (\#)         & 741           & 791           & 649          & 483           & 553           & 698           & 741           & 574           & 297           & 600 & -0.02 &  1.00         \\
\hline
domain-dependent (F1)   & 0.58          & 0.61          & 0.71         & 0.54          & 0.49          & 0.46          & 0.64          & 0.61          & 0.31          & 0.55 & 0.20 & 0.70         \\
domain-independent (F1)   & 0.68          & 0.66          & 0.71         & 0.64          & 0.65          & 0.63          & 0.71          & 0.69          & 0.48          & 0.61 & 0.28 & 0.76         \\
AL-trained (F1)   & 0.65	    & 0.67	        & 0.74         & 0.65	       & 0.62	    & 0.63          & 0.72         & 0.69	         & 0.50	        & 0.60& 0.23 & 0.68          \\
\hline
domain-dependent (std)  & 0.06         & 0.06         & 0.09         & 0.08         & 0.05         & 0.06         & 0.04         & 0.11         & 0.06         & 0.07 & 0.29 & 0.28      \\
domain-independent (std) & 0.04 & 0.04 & 0.11 & 0.08 & 0.07 & 0.05  & 0.03 & 0.04 & 0.06 & 0.03 & -0.11 & -0.05 \\
AL-trained (std) & 0.04    & 0.04	& 0.09	& 0.08	& 0.07	& 0.04	& 0.07	& 0.05	& 0.15	& 0.02 & -0.41 & -0.72 \\
\end{tabular}
\label{tab:correlation_variables}
\end{table}

Table \ref{tab:correlation_variables} summarises the results of our correlation analysis. The active learning classifier (AL-trained) has been trained with 52 \% training data sampled by MNLP.
For the domain-dependent, domain-independent and AL-trained classifier we observe a strong correlation between F1 and number of concepts per domain (\textit{R} 0.70, 0.76, 0.68) and a weak correlation between $\kappa$ and F1 (\textit{R} 0.20, 0.28, 0.23). Thus, we can hypothesise that the number of annotated concepts in a particular domain has more influence on the performance than the inter-annotator agreement.

The correlation values for std is different between the classifier types.
For the domain-dependent classifier the correlation between $\kappa$ and std (\textit{R} 0.29), and the number of concepts per domain and std (\textit{R} 0.28) is slightly positive. In other words: the higher the agreement and the size of the domain, the higher the variance of the domain-dependent classifier. This is different for the domain-independent classifier as there is no correlation anymore. For the AL-trained classifier there is, on the other hand, a moderate negative correlation between $\kappa$ and std (\textit{R} -0.41), and a strong negative correlation between number of concepts per domain and std (\textit{R} -0.72), i.e. higher agreement and larger amount of training data in a domain lead to less variance for the AL-trained classifier.
We hypothesise that more diversity through several domains in the domain-independent and the AL-trained classifier leads to better performance and lower variance by introducing an inductive bias.

\section{Conclusions}
In this paper, we have introduced the novel task of \textit{domain-independent concept extraction} from scientific texts. During a systematic annotation procedure involving domain experts, we have identified four general core concepts that are relevant across the domains of Science, Technology and Medicine. To enable and foster research on these topics, we have annotated a corpus for the domains.
We have verified the adequacy of the concepts by evaluating the human annotator agreement for our broad STM domain corpus.
The results indicate that the identification of the \textit{generic} concepts in a corpus covering 10 different scholarly domains is feasible by non-experts with moderate agreement and after consultation of domain experts with substantial agreement (0.76 $\kappa$).

We have presented two deep learning systems which achieved a fairly high F1 score (65.5\% overall). The domain-independent system noticeably outperforms the domain-dependent systems, which indicates that the model can generalise well across domains. We also observed a strong correlation between the number of annotated concepts per domain and classifier performance, and only a weak correlation between inter-annotator agreement per domain and the performance. We can hypothesise that more annotated data positively influence the performance in the respective domain.

Furthermore, we have suggested active learning for our novel task. We have shown that only approx. 5 annotated abstracts per domain serving as training data are sufficient to build a performant model. Our active learning results for SciERC~\cite{Luan2018MultiTaskIO} and ScienceIE17~\cite{augenstein2017semeval} datasets were similar.
The promising results suggest that we do not need a large annotated dataset for scientific information extraction. Active learning can significantly save annotation costs and enable fast adaptation to new domains.

We make our annotated corpus, a silver-labelled corpus with 62K abstracts comprising 24 domains, and source code publicly available\footnote{https://gitlab.com/TIBHannover/orkg/orkg-nlp/tree/master/STM-corpus}. We hope to facilitate the research on that task and several applications, e.g. academic search engines or research paper recommendation systems.
 
In the future, we plan to extend and refine the concepts for certain domains.
Besides, we want to apply and evaluate the information extraction system to populate a research knowledge graph.
For that we plan to extend the corpus with co-reference annotations~\cite{Lee2017EndtoendNC} so that mentions referring to the same concept can be collapsed.

%
%
%
\bibliographystyle{splncs04}
\bibliography{emnlp-ijcnlp-2019}

\end{document}